\documentclass [twocolumn,showpacs,amsymb,amsmath,nobibnotes,superscriptaddress,aps,prl]{revtex4}

 \usepackage {bm}
 \usepackage{graphicx}
 \usepackage{xcolor}
 \usepackage{amsmath}
\begin{document}

\title{Triplet pairing driven by Hund's coupling in doped monolayer MoS$_2$}
\author{Jie Yuan$^{*1}$, Carsten Honerkamp$^{\dag}$}
\affiliation{Institute for Theoretical Solid State Physics, RWTH Aachen, 52056, Germany \\ and JARA-FIT}

\begin{abstract}
We investigate superconducting pairing driven by electron-electron interactions in a theoretical model for monolayer MoS$_2$ with the temperature-flow functional renormalization group(fRG). At low doping, the dominant instability is toward odd-parity pairing with $f$-wave Mo-nearest-neighbor structure. We compute the fRG phase diagram versus electron doping below the van Hove filling of the conduction band.  In the superconducting regime, the critical temperature grows with doping, comparable to the experiments. Near van Hove filling the system favors a ferromagnetic state. We demonstrate that the triplet pairing is driven by ferromagnetic fluctuations and that the multiorbital nature of the conduction band as well as the Hund's coupling appear crucial in making the physics of MoS$_2$ different from e.g. doped graphene. 
\end{abstract}

\pacs{71.10.Fd, 74.20. Mn, 74.20.Rp, 74.70.-b}

\maketitle
Superconductivity with critical temperatures around 10K has been reported in MoS$_2$ flakes by electron doping using combined ionic-liquid and solid-state gating\cite{exp,exp1}. The observed superconducting dome as function of doping suggests the possibility of unconventional pairing. As Mo is a transition metal and the electronic bands near the Fermi level in MoS$_2$ have predominant $d$-orbital weight, electron-electron interactions are likely to play a role.

Monolayer  MoS$_2$ in the 2H-structure can be regarded as a sibling of of graphene. Compared to graphene with its two C sublattices, in MoS$_2$, one sublattice site is occupied by Mo, while the other sublattice  is formed by two S atoms displaced above and below the Mo plane. 
The layered 2H-structure is a direct gap semiconductor with gap size $\sim 1.8$eV\cite{dft1,dft2}. Bulk MoS$_2$ has an indirect gap\cite{dft3}. Upon electron doping, the Mo-$d$-dominated conduction band develops pockets around the $K$ and $K'$-points. By continued doping, these pockets merge with small extra pockets around the $M$-points. At the merging points, the density of states (DOS) exhibits van-Hove singularities. 

The possible nature of superconductivity in MoS$_2$ is of high interest to theory\cite{sc1,sc2,sc3,sc4,sc5}. The work by Roldan et al.\cite{sc1} compares pairing due to electron-electron interactions and mediated by phonons using effective intra- and inter-pocket interactions between the pockets that form in the conductions band around the $K$ and $K'$ points upon doping. Their conclusion is that phonon-mediated interactions have a smaller effect, and that the resulting pairing is driven by the short-ranged Coulomb interactions, and comes out as odd parity. Topological $p$-wave states due to electronic interactions are also reported by a mean-field theory\cite{sc5}. A purely  phonon-mediated mechanism based on first principles is proposed  in Refs. \onlinecite{sc2,sc3,sc4}.

The purpose of this paper is extend the study of Ref. \onlinecite{sc1} and to explore possible ground states of MoS$_2$ using a purely electronic model.  
We apply a temperature-flow functional renormalization group (fRG) technique\cite{tfrg,frgrev} to a three-orbital tight-binding model for MoS$_2$ proposed in Ref. \onlinecite{dft2}. The $T$-flow fRG is an unbiased approach and sufficiently flexible to treat the doping-dependent Fermi surfaces with enough angular resolution such to disentangle different pairing types. Importantly in our context, it is not blind with respect to ferromagnetic fluctuations\cite{tfrg}. 
Furthermore, we use as input effective interaction parameters obtained by constrained random-phase-approximation (cRPA) calculations. Hence, in addition to confirming the qualitative picture for the paired state of Ref. \onlinecite{sc1}, we give an ab-initio-based estimate of the critical temperatures as function of the doping, unveil the origin of electronic pairing as mediated by ferromagnetic fluctuations and the importance of Hund's rule coupling, and give an overview of the competing phases. This suggests experimental searches for ferromagnetic correlations in MoS$_2$ and related systems.


\begin{figure}
\centering
\includegraphics[width=8cm]{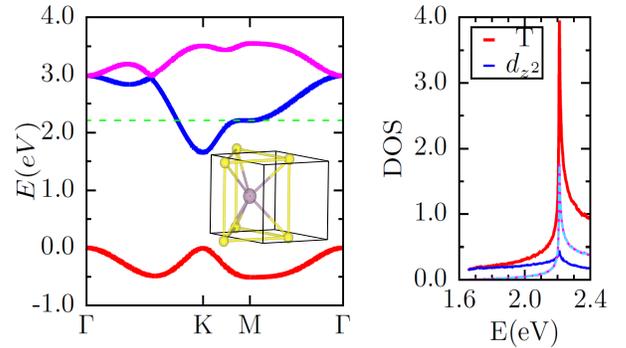}
\caption{{Left:} Band structure along the contour $\Gamma - K-M-\Gamma$.  {Inset:} Schematic view of crystal structure of monolayer MoS$_2$. The purple sphere represents the Mo, the yellow spheres sulfur atoms. {Right:} Electronic density of states (DOS) of the 3-band model. The red line shows the total DOS with the van-Hove singularity at $E=2.2$eV. The blue line is the band-projected DOS for the $d_z^2$-orbital, while $d_{x^2-y^2}$- and $d_{xy}$-contributions are shown by the coinciding solid purple and green dashed lines, respectively. }
\label{fig1}
\end{figure}

For the low-energy physics in MoS$_2$ we use a three-band tight-binding model from Ref. \onlinecite{dft2}. In the localized state basis 
$\left(
\begin{array}{ccc}
d_{z^2}^{\dag} , &d_{x^2-y^2}^{\dag} , &d_{xy}^{\dag}
\end{array}
\right)$
the tight-binding model Hamiltonian is given by a $3\times 3$ matrix. For simplicity, we only consider nearest-neighboring hopping terms with the parameters of Ref. \cite{dft2}, resulting in bands as plotted in Fig.~1(a). 
Spin-orbit coupling will be discussed more below. Doped electrons enter first into the middle band around $\mathbf K(\mathbf K')$, forming Fermi pockets. In Fig.1(b), we plot the density of states (DOS) of this band. It exhibits a van Hove (VH) singularity around $E=2.21$eV, which is when the extending Fermi pockets touch developing small pockets around $M$. In this paper, we focus on smaller doping levels $x$ up to van-Hove filling. Around $K$, $K'$, the doped electrons have predominant  $d_{z^2}$ character, while near the VH level, the $d_{x^2-y^2}$ and $d_{xy}$ components are stronger, as can be seen from Fig.~1(b). 
In the DFT band structures and more elaborate tight-binding parametrizations\cite{dft2}, additional band minima along the $\Gamma$-$K$ direction not captured in our simplified model can lead to additional Fermi pockets at higher $x$. These however have smaller DOS than e.g. the $M$-point regions and should therefore be of minor importance for the trends found in this work.
The interactions are given by
\begin{equation}
\begin{aligned}
H_{\texttt{int}} &=U_{00} \sum _{i} 
 \sum _{a} n_{ia\uparrow} n_{ia\downarrow} + U_{00}' \sum_i\sum_{a<b} n_{ia}n_{ib}\\
&+ J_H\sum_i \sum _{a<b,s,s'} \psi^{\dag} _{ias} \psi^{\dag} _{ibs'} \psi_{ias'} \psi_{ibs}\\
&+J_H \sum_i \sum_{a<b} \left( \psi_{ia\uparrow} ^{\dag} \psi_{ia\downarrow} ^{\dag} \psi_{ib\downarrow} \psi_{ib\uparrow} + H.c.\right) 
\end{aligned}
\end{equation}
with $U_{00}=2.61$eV, $U_{00}'=2.09$eV, $J_H=0.27$eV, computed from first principles by cRPA \cite{crpa,crpa1,crpa2}. 
In a broader view, besides the minor splitting of the van Hove singularities near the $M$-points, 
the Fermi surface evolution in our model for MoS$_2$ is reminiscent of that in doped graphene (e.g. studied in \cite{dwave,kiesel,wang3}). The natural question arises if the phenomenology is similar. We we argue that the multi-orbital nature due to the mixing of 3 $d$-orbitals at the Fermi level in the conduction band causes qualitative differences, at least for model parameters close to the ab-initio predictions. We in fact show that the Hund's coupling is a tuning parameter that takes us from the 'doped-graphene' scenario to  MoS$_2$.

\begin{figure}
\centering
\includegraphics[width=8.6cm]{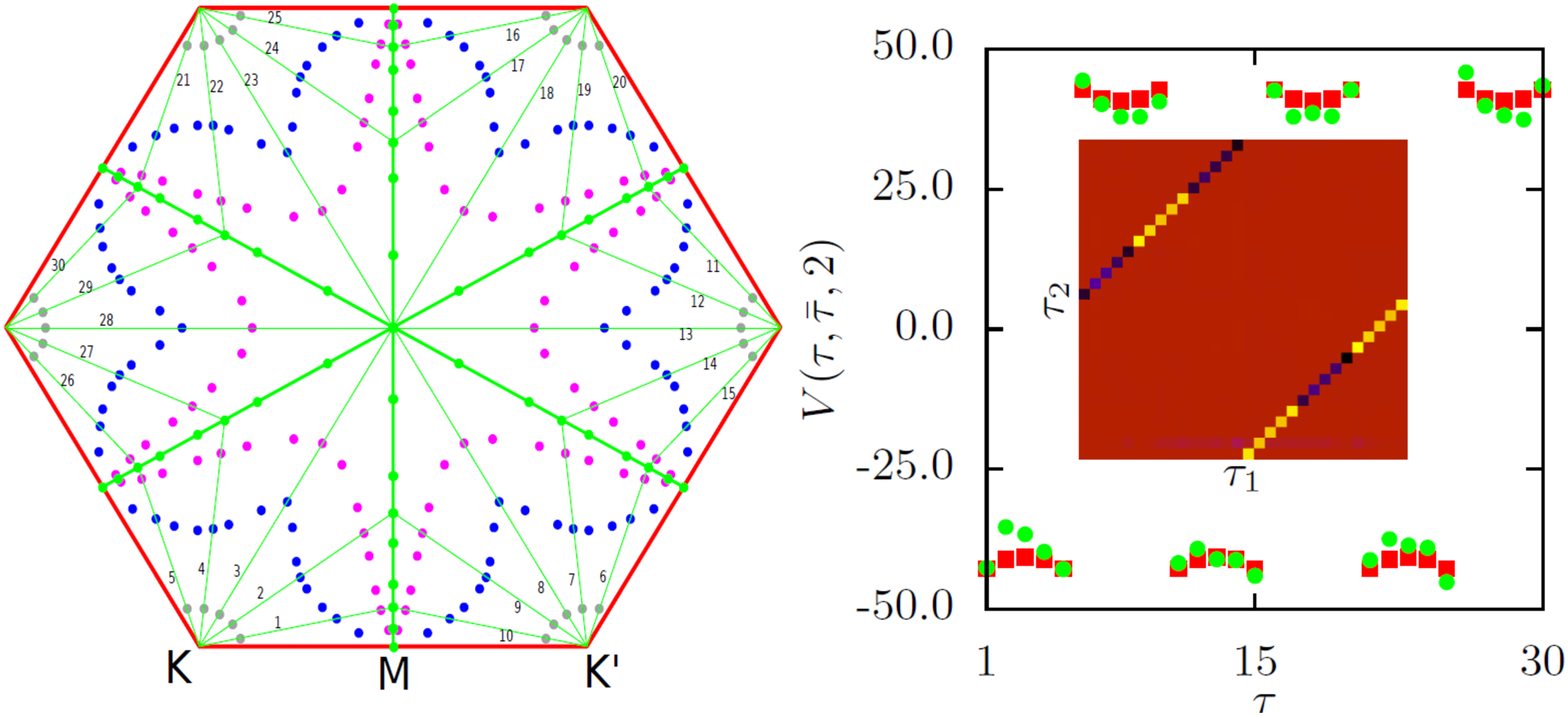}
\caption{Left: Patching scheme used in the fRG. At low doping $x$ the Fermi surface (FS) is around the $K$, $K'$ points. On the FS we define 30 patch points (grey filled circles).  The patch index $\tau$ runs from $1$ to $30$.  In addition to these FS points, we define two other sets of patch points (blue and pinks circles). This way, the energy dispersion is monotonic in each sector in radial direction. 
 Right: Pair scattering at small $x$. The green symbols in the main plot show the fRG data for the low-T pair scattering  $V(\tau,\bar{\tau},2)$ near the instability, where the two incoming wavevectors belonging to index $\tau$ and $\bar\tau$ add up to total zero momentum. The first outgoing wavevector is fixed at patch point 2. The red square symbols show a fit to the fRG data using a nearest-neighbor-(i.e. Mo-Mo-)bond pairing $-f_p(\mathbf k[\tau]) f_p^*(\mathbf k[2])$. {Inset:} $V(\tau_1,\tau_2,\tau_3)$ as function of patch indices $\tau_1$ and $\tau_2$ for $\tau_3=2$. Clearly visible is that only interactions with a special relation between $\tau_1$ and $\tau_2$ (implying zero total incoming wavevector) become large. For the data in this plot, $\mu=1.8$eV.   }
\label{fig2}
\end{figure}

We use the temperature-flow fRG\cite{tfrg,frgrev} to analyze the possible instabilities due to interactions. The main idea of the method is that at very high temperatures $T$ the interactions of the model are given by the bare ones, as all loop corrections are killed by $T$. Upon lowering $T$, the $T$-flow tracks the build-up of the loop corrections. The main object of study is an interaction function $V_T (k_1,k_2,k_3) $ that depends in our case on three wavevectors $\vec{k}_1$, $\vec{k}_2$ and $\vec{k}_3$ for two incoming legs 1 and 2 and one outgoing leg 3 (with the same spin projection as leg 1) in the conduction band. The flow equations for  $V_T (k_1,k_2,k_3)$ for charge and spin-rotational invariance are taken from the general fRG formalism for one-particle irreducible vertex functions (for a review, see Refs. \cite{frgrev,plattrev} or also Ref. \onlinecite{tfrg}) and are given in the supplementary material.
We start the flow at an initial $T$ of roughly twenty times the bandwidth. Lowering $T$ generally leads to a strong differentiation of the components of $V_T (k_1,k_2,k_3)$, where certain components become large at low $T$ and actually diverge. These generalized Cooper instabilities can then be interpreted as ordering tendency in a particular channel, depending on which combination of $k_1$, $k_2$ etc. actually grows most strongly.  
The flow is stopped when the maximal component of $V_T$ grows larger than thirty times larger than the bandwidth. This defines a critical temperature $T_c$ which should be understood as upper estimate for any true ordering. More information about the interpretation of the fRG flow can e.g. be found in the reviews of Refs. \cite{frgrev} and \cite{plattrev}. In a true two-dimensional system, collective effects should in fact destroy any long range order that breaks continuous symmetries. These effects are however not included in the used approximation in the fRG, so do not play a role in our results. Also in view of the embedding of the experimental systems in three dimensions we do not expect that such collective effects have to be considered at this stage.
 Further, we  patch Brillouin zone around the $K$, $K'$ points with patch center points numbered by $\tau =1 \dots, N$, as seen in Fig. \ref{fig2}. Then we discretize the coupling function as $V_T(\tau_1,\tau_2,\tau_3)$, i.e. with $N^3$ components.  At each corner $K$, $K'$, we take 5 patches (schemes with more patches have also been checked, the result is stable). We divide the radial direction of each patch into three sectors, providing some more  resolution in the radial direction. In the loop calculations, we use 3 radial integration lines in each patch, but did also checks with more lines. 

We run the $T$-flow fRG for a wider range of electron dopings $x$. For small $x$, we find pairing instabilities with total-momentum-zero Cooper pairs in the odd-parity channel, while at larger $x$ near van Hove filling, a ferromagnetic (FM) instability is found. The latter can be understood as a renormalized version of a Stoner instability due to a large DOS at the Fermi level.  

The data for a prototypical pairing instability at $x=0.03$ is shown in Fig. \ref{fig2} on the right side. The inset shows the coupling function for incoming and outgoing legs at the Fermi surface, as function of the two incoming patch indices, with the first outgoing leg fixed in patch 2. The two diagonal features are exactly on the lines where the two incoming wavevectors add up to zero. These pair-scattering processes grow to very large absolute values and cause the instability. We also see the sign structure with pieces of large positive and large negative values. This is detailed more in the main part of Fig. \ref{fig2}, right side, which is basically a cut along the zero-total-momentum diagonal features in the inset. By looking up the patch numbers Fig. 1 one extracts that the pairing changes sign upon rotating the incoming pair by 60 degrees, from one Fermi pocket around, e.g. around $K$, to another pocket, e.g. around $K'$. The same sign change is found when the incoming pair is fixed and the outgoing pair is varied. The couplings do not vary much around $\mathbf K(\mathbf K')$ points, i.e. is no nodal structure along the Fermi pockets. 
This odd-parity pairing can be identified to transform according to the one-dimensional $A'_2$ irreducible representation of $D_{3h}$ point group of MoS$_2$. In real space it can be understood as having its main contributions on the bonds from one Mo site to its 6 nearest neighbor Mo sites, with the sign alternating upon 60 degrees rotation, implying the odd parity. The form factor in wave vector space can also be interpreted as $f$-wave and reads
$f_p(\mathbf k)= \sin \left( \sqrt 3 k_x\right)-2 \sin \left( \frac{\sqrt 3k_x }{2}\right) \cos \left(\frac{3k_y}{2} \right)$.
The fit to this form factor shows good agreement (red squares and green circles in Fig.3).

We also run a momentum-cutoff fRG flow\cite{frgrev} which disfavors FM fluctuations with small wave-vector transfer across the Fermi surface down to very low scales\cite{tfrg}. These flows do not give any instabilities but show some tendency to even-parity pairing. Therefore, we can conclude that $f$-wave pairing  is driven by $\mathbf Q =\mathbf 0$ ferromagnetic fluctuations.  A similar conclusion is reached by turning off specific channel spin fluctuation channels in the flow equation directly.  

\begin{figure}
\centering
\includegraphics[width=8cm]{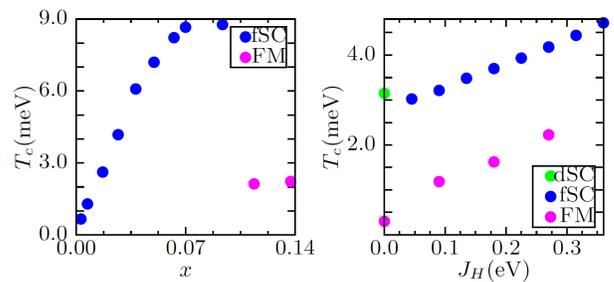}
\caption{Left: fRG phase diagram of monolayer MoS$_2$ versus electron doping $x$. The blue dots give the $T_c$s according to fRG for the $f$-wave superconducting phase (\textbf{SC}). The purple dots represent the $T_c$s ferromagnetic state (\textbf{FM}), now scaled down by a factor of 100. The fRG $T_c$s should be interpreted as upper bounds for true ordering temperatures. Right: fRG critical temperatures as function of Hund's coupling $J_H$ for both the triplet pairing (blue circles,
) and FM instability (pink circles, 
 again multiplied by 0.01). Notably, for $J_H\approx 0$ the dominant pairing instability appears in the $d$-wave singlet pairing channel. }
\label{fig4}
\end{figure}

In Fig. \ref{fig4}  we show the fRG phase diagram of our MoS$_2$ model with doping. In the low doping regime, the discussed $f$-wave superconductivity is the dominant instability, with a relatively low critical temperature $T_c$ (of order $1.0$meV). We can also see that the critical temperature grows with increasing electron doping.   The highest scale for pairing for our parameters is around $8.0$meV, considerably higher than the experimental data around $1.3$meV. We comment on this further below. The optimal doping for the $f$-wave pairing is reached around $x\sim 0.1$. This shows good agreement with experiment\cite{exp}. 
The increasing DOS with growing $x$ also enhances the FM fluctuations. In simple mean-field or Stoner pictures the critical interaction strength required for FM order is inversely proportional to the DOS at the Fermi level, and as the DOS diverges at the van Hove filling, one will definitely find ordering in these approaches.  In the fRG calculations we also obtain a regime where the leading instability is of FM type, however in reduced extent compared to mean-field theory due to the competition with other fluctuations. Hence, the prediction form the fRG approach would be that there are predominant FM fluctuations around the van Hove filling. Whether the ordering actually occurs or not cannot be answered by our calculations. For the Hubbard model on the square lattice at large second-nearest-neighbor hopping, the fRG also predicts FM ordering, in agreement with variational Montecarlo results\cite{hlubina}. However, the van Hove filling may be hard to realize precisely in the experiment as disorder and spatial fluctuations will tend to smear out the diverging DOS.  Nevertheless, the FM correlations should still be visibly enhanced.

At small $x$ the states near the Fermi level are mainly of $d_{z^2}$-type. Thus the local intraorbital repulsion, $U_{00}$, should be important. We hence run the flow with only $U_{00}$ being nonzero. As expected, the flow still ends up with the $f$-wave instability.  However, $T_c$ is lowered from about 4.8meV with full interactions down to $1.2$meV.  This shows that interorbital processes contribute significantly to the pairing.  Next we vary the Hund's coupling $J_H$ in the interactions as given in Eq. (3), with $U_{00}$ and $U_{00}'$ fixed, and the doping  fixed at $\mu =1.8$eV. From Fig.~4,  we can read off an enhancement of $T_c$ in the $f$-wave regime by a larger $J_H$, from $3.0$meV to around $8.0$meV.  Thus, $J_H$ enhances the $f$-wave instability as well.  Even more strikingly, when we set the $J_H =0.0$eV, the system gives way to an even-parity pairing structure, with most attractive eigenvalues in the $d$-wave symmetry channel\cite{dwave}, associated with the two-dimensional $E'$ irreducible representation of $D_{3h}$ point group. Complex combinations of $d_{xy}$ and $d_{x^2-y^2}$ break time-reversal symmetry, leading to chiral edge modes as known from strongly doped graphene models\cite{dwave,chiral}. 
We also study the importance of the Hund's coupling $J_H$  in the FM regime. The data is shown as pink dots in Fig.~5. The critical temperature is again enhanced dramatically when $J_H$ is increased. The Hund's coupling intensifies the FM fluctuations just as it strengthens the $f$-wave pairing. Such an effect of $J_H$ also reported in models for NaCo$O_2\cdot yH_2$O\cite{hund}.  We also observe a kink in the critical temperature as a function of $J_H$ at $J_H \approx 0.03$eV where the system undergoes a phase transition from even-parity pairing at small $J_H$ to odd-parity pairing. Hence $J_H$ has a key role as a switch between two different situations.
On the one hand, we have the single-orbital scenario of graphene, where $J_H$ is ineffective as the bands near the Fermi level are just made from C $p_z$-states. Then one finds strong antiferromagnetic fluctuations and chiral $d$-wave singlet pairing. For MoS$_2$, the multi-orbital character of the conduction band makes $J_H$ important and, for realistic values of $J_H$ according to cRPA, one finds predominant FM fluctuations and triplet pairing. 

These considerations indicate that the superconducting ground state is quite susceptible to parameter changes. This is one reason why the rather large pairing scales $\sim 10$meV exceeding the experimental scales by one order of magnitude should not be taken as a failure of this theoretical picture. Some reduction of the $T_c$s should be expected if we would be able to include self-energy corrections or more collective fluctuations into the fRG, which is however not yet technically feasible for multiband systems. In view of the importance of the Hund's rule coupling, we refer to recent works by Georges et al.\cite{hundmetal}, who pointed out that $J_H$ typically degrades the quasiparticle weight resulting in a bad {\em Hund's metal}. This may be another important factor that reduces the superconducting $T_c$s compared to our study.
Then there are uncertainties about the actual experimental system. Disorder is expected to have an effect on the $T_c$ of any unconventional superconductor, although in the $f$-wave case only large-momentum impurity scattering will have an impact of the phase structure of the pairing.  On the positive side, the high scales from the theory might actually show that there may be some headroom for reaching higher $T_c$s in Mo$S_2$ and related systems.

In MoS$_2$ and WS$_2$,  the spin-orbit coupling splitting can reach the order $0.1$eV \cite{dft1,dft2}. In the conduction band near $K$, $K'$, the actual values may be smaller $\sim 3$meV\cite{burkard}. The $S_z$-component is conserved around the small Fermi pockets shown in Fig. 2 which then split into two pockets around each $K$, $K'$. By time-reversal symmetry, the pockets with the same radius at different pockets will have opposite spin projections. The $f$-wave pairing does not violate time-reversal symmetry\cite{pairing} and should not be changed strongly by the small splitting near $K$, $K'$.  In such a case, the $z$ component of the gap function $d_z(\mathbf k)$ can be expected to be dominant. This corresponds to $S_z=0$ pairing.

In conclusion, we investigate the possible pairing structure in monolayer-MoS$_2$ by the unbiased $T$-flow fRG. For interaction parameters obtained by cRPA, the leading instability at low electron doping lies in the $f$-wave pairing channel. We argued that ferromagnetic fluctuations provide the pairing glue.  The phase diagram exhibits dominant $f$-wave pairing up to  $x\sim0.10$. For doping close to the van-Hove filling, sufficiently strong interactions result in  FM order. The pairing $T_c$ is maximal around an optimal doping of $x \sim 0.1$, in quite a good agreement with experiment. Varying the Hund coupling strength $J_H$ showed strong effects on both the $f$-wave pairing scale and the FM tendencies. Notably, when $J_H$ is relatively small or set to zero, the leading pairing instability is in the singlet $d$-wave channel. This emphasizes the importance of the Hund's coupling and the multi-orbital nature of the Mo-$d$-dominated conduction band.

Our study extends earlier work by Roldan et al.\cite{sc1} who already found $f$-wave pairing. It clarifies the nature of the pairing mechanism, the influence of the model parameters, the importance of FM fluctuations and the Hund's coupling. Future studies should  include the phonon-mediated electronic interactions. This should allow one to compare the relative importances of these two distinct pictures for the superconductivity in MoS$_2$.

We acknowledge useful discussions with Ersoy \c{S}a\c{s}{\i}o\u{g}lu, Tim Wehling,  Stefan Haas and Fu-Chun Zhang and support by DFG-FOR 912 and SPP1459, as well as DFG-Ho2422/10-1.  

\clearpage
\linespread{1}
\setcounter{figure}{0}
\renewcommand{\figurename}{Supplementary Fig.}
\section*{Supplementary Material}
\subsection*{T-flow equations}
The temperature-flow fRG uses the physical temperature of the system as flow parameter. 
As in previous works using this method\cite{tfrg,frgrev,plattrev}, the flow of the self-energy is neglected and the flow equations are truncated after the four-point vertex. Furthermore, the interaction vertex is approximated to be frequency-independent, and the external frequencies are set to zero. The wavevector dependence is treated in the $N$-patch discretization\cite{tfrg,frgrev,plattrev}.

The initial condition for the flow at the initial $T$ larger than the bandwidth of the system is the bare action, as all perturbative corrections vanish at high enough $T$. Then the flow leads to lower $T$ and picks up the different one-loop corrections for the interactions, each in second order in the vertices at the given temperature.  More precisely, the flow equations read in this case\cite{tfrg}
\begin{equation}
\begin{aligned}
\frac{d}{dT} V_{T} =\Gamma_{pp,T}+\Gamma^d_{ph,T}+\Gamma_{ph,T}^c
\end{aligned}
\end{equation}
where explicitly
\begin{equation}
\begin{aligned}
\Gamma_{pp,T}&(k_1,k_2,k_3,k_4)=-\int dp \, L(p,k_1+k_2-p) \\
&\times V_{T} (k_1,k_2,p) 
 V_{T} (p,-p+p_1+p_2,p_3) 
\end{aligned}
\end{equation}
\begin{equation}
\begin{aligned}
\Gamma^d_{ph,T}& (k_1,k_2,k_3,k_4) =-\int dp  \, L(p,p+p_1-p_3) \\
&\times[ -2 V_{T}(k_1,p,k_3)V_{T}(p+k_1-k_3,k_2,p) \\
&+V_{T}(k_1,p,p+k_1-k_3) V_{T} ( p+k_1-k_3,k_2,p)\\
&+V_{T}(k_1,p,k_3) V_{T}(k_2,p+k_1-k_3,p)]
\end{aligned}
\end{equation}
\begin{equation}
\begin{aligned}
\Gamma^c_{ph,T}&(k_1,k_2,k_3,k_4)=-\int dp  \, L(p,p+p_2-p_3) \\
&V_{T} (k_1,p+k_2-k_3,p) V_{T} (p,k_2,k_3)
\end{aligned}
\end{equation}
The $k_i$ indices are patch numbers, corresponding to wavevectors in the conduction band.
In the temperature flow without self-energy corrections, the one-loop integrals are given by
\begin{equation}
L(p,p')=\frac{d}{dT} \left( \frac{1}{i\omega -\epsilon_{\mathbf p}}\frac{1}{i\omega' -\epsilon_{\mathbf p'}} \right) \, . 
\end{equation}
In most situations, the usual momentum-shell cutoff schemes which have a cutoff-derivative in the loop diagrams give equivalent flows as the TfRG. In the case of particle-hole fluctuations with small wavevector transfer, a momentum-shell cutoff would lead to underestimation of these processes at intermediate scales while the $T$-flow incorporates these contributions in the same way as fluctuations with larger wavevector transfers. Hence the $T$-flow can capture ferromagnetic instabilities and is hence a more unbiased fRG-method. Note that also frequency cutoff can be used to circumvent this shortcoming of the momentum-shell schemes\cite{husemann,giering,wang3}.

\end{document}